\documentclass[10pt]{article}
\usepackage{epsf}

\newlength{\dinwidth}
\newlength{\dinmargin}
\def\lapproxeq{\lower .7ex\hbox{$\;\stackrel{\textstyle <}{\sim}\;$}}
\def\gapproxeq{\lower .7ex\hbox{$\;\stackrel{\textstyle >}{\sim}\;$}}

\def\be{\begin{equation}}
\def\ee{\end{equation}}
\def\bea{\begin{eqnarray}}
\def\eea{\end{eqnarray}}

\def\bb{{b\bar{b}}}

\begin{document}
\titlepage

\begin{flushright}
IPPP/03/84 \\
DCPT/03/168\\
12 January 2004 \\
\end{flushright}

\vspace*{4cm}

\begin{center}
{\Large \bf Hunting a light CP-violating Higgs via diffraction at the LHC}

\vspace*{1cm} \textsc{V.A.~Khoze$^{a,b}$, A.D. Martin$^a$ and M.G. Ryskin$^{a,b}$} \\

\vspace*{0.5cm} $^a$ Department of Physics and Institute for
Particle Physics Phenomenology, \\
University of Durham, DH1 3LE, UK \\[0.5ex]
$^b$ Petersburg Nuclear Physics Institute, Gatchina,
St.~Petersburg, 188300, Russia \\
\end{center}

\vspace*{1cm}

\begin{abstract}
We study the central diffractive production of the (three neutral) Higgs bosons,
with a rapidity gap on either side,
  in an MSSM scenario with CP-violation.
   We consider the $b\bar{b}$ and $\tau\bar{\tau}$
  decay for the light $H_1$ boson and the four $b$-jet final state for the
  heavy $H_2$ and $H_3$ bosons, and discuss the corresponding backgrounds.
A direct indication of the existence of CP-violation can come from the observation
of either an azimuthal asymmetry in the angular
   distribution of the tagged forward protons (for the exclusive $pp\to p+H+p$
    process) or of a sin$2\varphi$ contribution in the azimuthal correlation between the
    transverse energy flows in the proton fragmentation regions for the
    process with the diffractive dissociation of both incoming protons
    ($pp\to X+H+Y$).
    We emphasise the advantage of reactions with the rapidity gaps
    (that is production by the pomeron-pomeron fusion) to probe CP parity and to
determine the quantum numbers of the produced central object.
\end{abstract}


\section{Introduction}

It is known that third generation squark loops can introduce sizeable CP violation in the Higgs potential of the
Minimal Supersymmetric Standard Model (MSSM), if the soft-supersymmetry-breaking mass parameters of the third
generation are complex; see, for example,~\cite{AP}.   As a result, the neutral Higgs bosons will mix to produce
three physical mass eigenstates with mixed CP parity, which we denote $H_1,H_2$ and $H_3$ in order of increasing
mass.  A benchmark scenario of maximal CP violation, called CPX, was introduced in Ref.~\cite{CEPW}. In this
scenario
\be |A_t|=|A_b|=2 M_{\rm SUSY},\quad |\mu|=4 M_{\rm SUSY}, \quad M_{\tilde{Q}_3,\tilde{U}_3,\tilde{D}_3}=M_{\rm
SUSY},\quad |M_3|=1~{\rm TeV},  \label{eq:jan5a} \ee
where $A_f$ are are the soft-supersymmetry-breaking trilinear parameters of the third generation squarks and $\mu$
is the  supersymmetric higgsino mass parameter.  The phenomenological consequences of this model may be quite
spectacular.  In particular, the $H_1ZZ$ coupling of the lightest Higgs boson can be significantly suppressed;
see, for example, ~\cite{CEPW} and references therein. In this case, it was shown that the LEP2 data do not
exclude the existence of a light Higgs boson with mass $M_H<60$~GeV (40~GeV) in the minimal SUSY model with
$\tan\beta\sim3$--4 (2--3) and CP-violating phase
\be \phi_{\rm CPX} \equiv {\rm arg}(\mu A_t) = {\rm arg}(\mu A_b) =
 {\rm arg}(\mu A_\tau)={\rm arg} (\mu m_{\tilde g}) =
90^\circ~(60^\circ). \label{eq:A1} \ee
Since the  $H_1$ couplings to the $W$ and $Z$ gauge bosons become rather small,
it would be hard to detect the light Higgs via the processes
$e^+e^- \to Z^\star\to ZH_i$ or $e^+e^- \to Z^\star\to H_iH_j$.

It is therefore interesting to consider the possibility of observing  a light Higgs
boson at the LHC or
Tevatron collider. However, in general, it will be hard to
observe a light Higgs at hadron colliders via the $\bb$
decay mode because, in particular, the transverse momenta of the
outgoing $b$ and $\bar b$ jets are not large. As a consequence
the signal is
swamped by the QCD $\bb$ background\footnote{The prospects for observing such a light Higgs
in conventional search channels, at the Tevatron and the LHC, were studied in~\cite{CHL,CEMPW}.}.
Therefore it was proposed~\cite{cox} to search
for a CP-violating
light Higgs boson in the {\em exclusive} process $pp\to p + H + p$ at
hadron colliders, where the $+$~signs denote the presence of large
rapidity gaps.  Over the past few years such exclusive diffractive processes
have been considered as a promising way to search for manifestations of
New Physics in high energy proton-proton collisions; see,
 for instance,~\cite{KMRcan,INC,cox,KKMRCentr,DKMOR,CR}.   These processes
have both unique experimental and theoretical advantages in hunting for Higgs bosons
as compared to the traditional non-diffractive approaches.  In particular,
in the exclusive diffractive reactions the
$\bb$ background
is suppressed~\cite{Liverpool,KMRItal,KMRmm,DKMOR}, and it may be feasible
to isolate the signal.

In the present paper we discuss the central
{\em exclusive} diffractive production (CEDP) in more detail.
We compare the signal and the background for
observing a light neutral Higgs boson via $H_1\to\bb$ and $H_1\to\tau\tau$
decay modes. Then we evaluate the asymmetry arising from the interference
of the P-even and P-odd production amplitudes. Note that this asymmetry is
the most direct manifestation of CP-violation in the Higgs sector.
Finally we consider the exclusive diffractive production of the heavier neutral Higgs bosons,
$H_2$ and $H_3$, followed by the decays $H_2$ or $H_3\to H_1H_1\to 4 b$-jets.\\

 For numerical estimates, we use the
formalism to describe central production in diffractive exclusive processes of~\cite{INC}, and the parameters
(that is the masses, width and couplings of the Higgs bosons) given by the code "CPsuperH" \cite{Lee}, where we
choose $\phi_{\rm CPX}=90^\circ$, $\rm tan\beta=4$, $M_{\rm SUSY}=0.5$~TeV, (that is $|A_f| = 1$~TeV, $|\mu| =
2$~TeV, $|M_{\tilde g}|=1$~TeV) and the charged Higgs boson mass $M_{H^\pm}=135.72$ GeV so that the mass of the
lightest Higgs boson, $H_1$, is $M_{H_1}=40$ GeV.\footnote{The values are chosen to provide an 'optimistic'
scenario for the observation of a CP-violating Higgs boson in CEDP.}

\begin{figure}[htb]
\centerline{\epsfxsize=4cm\epsfbox{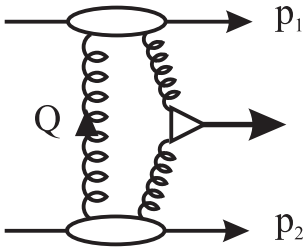}}%
\caption{Schematic diagram for the exclusive central production of a
light Higgs boson.}
\label{fig:1}
\end{figure}

The exclusive process is shown schematically in
Fig.~\ref{fig:1}.
The cross section may be written\cite{INC} as the product of the effective gluon--gluon luminosity ${\cal L}$, and
the square of the matrix element of the subprocess $gg\to H$.  Note that the hard subprocess is mediated by the
quark/squark triangles. For a CP-violating Higgs, there are two different vertices of the Higgs--quark
interaction: the scalar Yukawa vertex and the vertex containing the $\gamma_5$ Dirac matrix. Therefore the $gg\to
H$ matrix element contains two terms:\footnote{For calculations of $g_S$ and $g_P$ in the MSSM with CP-violation
see, for example, \cite{DMCL}.}
\be {\cal M}  =  g_S\cdot (e_1^\perp\cdot e_2^\perp) - g_P \cdot\varepsilon^{\mu\nu\alpha\beta}
e_{1\mu}e_{2\nu}p_{1\alpha}p_{2\beta}/(p_1\cdot p_2) \label{eq:1}
\ee
where $e^\perp$ are the gluon polarisation vectors and
$\varepsilon^{\mu\nu\alpha\beta}$ is the antisymmetric
tensor. In (\ref{eq:1}) we have used a simplified form of the matrix element which already accounts for gauge
invariance, assuming that the gluon virtualities are small in comparison with the Higgs mass. In forward
exclusive central production, the incoming gluon polarisations are
correlated, in such a way that the effective luminosity satisfies the P-even,
 $J_z=0$ selection rule~\cite{INC,KMRmm}.
Therefore only the first term contributes to the strictly forward cross
section. However, at non-zero transverse momenta of the recoil protons,
$p_{1,2}^\perp\neq0$, there is an admixture of the
 P-odd $J_z=0$ amplitude
of order $p_1^\perp p_2^\perp / Q_\perp^2$, on account of the $g_P$
term becoming active. Thus we consider non-zero recoil proton
transverse momenta, and demonstrate that the interference between the
CP-even ($g_S$) and CP-odd ($g_P$) terms leads to left-right asymmetry
in the azimuthal distribution of the outgoing protons. First, we
consider the background. Unfortunately, even in the exclusive process,
we show below that the QCD $\bb$ background is too large. However, we
shall see that it may be possible to observe such a CP-violating light
Higgs boson in the $H\to \tau\tau$ decay mode, where the QED background
can be suppressed by selecting events with relatively large outgoing
proton transverse momenta, say, $p_{1,2}^\perp>300$~MeV.

\section{ Exclusive diffractive $H_1$ production followed by $\bb$ decay }
First, we consider the exclusive double-diffractive process
\be pp\to p+(H\to\bb)+p
\label{eq:A2}
\ee
The signal-to-background ratio is given by the ratio of the cross
sections for the hard subprocesses, since the effective gluon--gluon luminosity ${\cal L}$ cancels out.  The cross
section for the $gg\to H$ subprocess\footnote{In \cite{INC} we denoted the initial state by $gg^{PP}$ to indicate
that each of the incoming gluons belongs to colour-singlet Pomeron exchange. Here this notation is assumed to be
implicit.}~\cite{INC}
\be \hat \sigma (gg\to H)\ =\ \frac{2\pi^2\Gamma(H\to gg)}{M_H^3}\delta\left(1-\frac{M_\bb^2}{M_H^2}\right)\ \sim\
{\rm constant}\times\delta\left(1-\frac{M_\bb^2}{M_H^2}\right), \label{eq:2}\ee
as the width\footnote{Strictly speaking, we should consider CP-even and CP-odd contributions to the width
separately, but it does not change the conclusion qualitatively.}, $\Gamma(H\to gg)$,
behaves as $\Gamma\sim \alpha_S^2 G_F M_H^3$,
 where $G_F$ is the Fermi
constant.  On the other hand, at leading order, the QCD background is
given by the $gg\to \bb$ subprocess %
\be
\frac{d\hat\sigma_{\rm QCD}}{dE_T^2} \sim \frac{m_b^2}{E_T^2}
\frac{\alpha_S^2}{M_\bb^2E_T^2},
\label{eq:3}
\ee
where $E_T$ is the transverse energy of the $b$ and $\bar b$ jets.  At leading order (LO), the cross section is
suppressed by the $J_z=0$ selection rule (which gives rise to the $m_b^2/E_T^2$ factor) in comparison with the
inclusive process.  The extra factor was crucial to suppress the background.  It was shown in~\cite{DKMOR} that it
is possible to achieve a signal-to-background ratio of about 3 for the detection of a Standard Model Higgs with
mass $M_H\sim 120$~GeV, by selecting $\bb$ exclusive events where the polar angle $\theta$ between the outgoing
jets lies in the interval $60^\circ<\theta<120^\circ$ if the missing mass resolution $\Delta m_{\rm missing} =
1$~GeV. The situation is much worse for a light Higgs, since the signal-to-background ratio behaves as
\be
\frac{\displaystyle \int\frac{d{\cal L}}{d\ln M_\bb^2}\ \hat\sigma(gg\to H)\ d\ln M_\bb^2}{\displaystyle
\int\frac{d{\cal L}}{d\ln M_\bb^2}\ \hat\sigma_{\rm QCD}\ d\ln M_\bb^2} \ \simeq \ \frac{G_F^2}{\displaystyle
\left(\frac{m_b^2}{M_\bb^2}\frac{1}{M_\bb^2}\right)\frac{2\Delta M_\bb}{M_\bb}} \ \sim \ M_\bb^5
\label{eq:4}
\ee
where we have used $\Delta\ln M_\bb^2 = 2\Delta M_\bb/M_\bb$. The $M^5$ behaviour comes just from dimensional
counting. As the experimental resolution $\Delta M_\bb$ is larger than the width of the Higgs, $\Gamma_H$, the
Higgs cross section (in the numerator) is driven by $G_F^2$, while the QCD background is proportional to $m_b^2$
and the size of the $\Delta M_\bb$ interval. To restore the dimensions we have to divide $m_b^2\Delta M_\bb$ by
$M_\bb^5$.  Thus, in going from $M_H\sim 120$~GeV to $M_H\sim40$~GeV, the expected leading-order QCD $\bb$
background increases by a factor of 240 in comparison with that
for $M_\bb=120$ GeV.

Strictly speaking, there are other sources of background~\cite{DKMOR}. There is the possibility of the gluon jet
being misidentified as either a $b$ or a $\bar b$ jet, or a contribution from the NLO $gg\to \bb g$ subprocess,
where the extra gluon is not separated from either a $b$ or a $\bar b$ jet. These contributions have no
$m_b^2/M_\bb^2$ suppression, and hence increase only as $M_H^{-3}$, and not as $M_H^{-5}$, with decreasing $M_H$.
For $M_H\sim 120$~GeV, the LO $\bb$ QCD production was only about 30\% of the total background. However, for
$M_{H_1}\sim40$~GeV, the LO $\bb$ contribution dominates. Finally, with the cuts of
Ref.~\cite{DKMOR}, we predict that the cross section of the $H_1$ signal
is\footnote{Note that our CEDP cross section is about two times larger than
that quoted in \cite{cox}. This difference occurs mainly because
we use an improved approximation for the unintegrated gluon densities.
To be specific, we use eq.(26) of \cite{MR01}, rather than
the simplified formula (4) of Ref.\cite{DKMOR} used in \cite{cox}. In addition we
allow for the transverse momenta $p^\perp_{1,2}$ of the recoil
protons in the gluon loop of Fig.1. For smaller boson masses,
$M_H\sim 40$ GeV, this leads to a steeper $p^\perp_{1,2}$
dependence
of the amplitude, which emphasizes larger values of the impact parameter, $b_\perp$,
where the absorptive effects are weaker. Therefore we obtain a
larger soft survival factor, $S^2\simeq 0.029$, at the LHC energy.
However, recall that a factor of 2
difference is within the accuracy of the approach\cite{DKMOR,KKMRCentr}.}
$$ \sigma^{\rm CEDP}(pp\to p+(H_1\to\bb)+p)\simeq 14~{\rm fb}$$
as compared to the QCD background cross section, with the same cuts\footnote{Here and in what follows we assume
that the proton and $b$-tagging efficiencies and the missing mass resolution in the case of a light Higgs boson
are the same as for the case of $M_{\rm Higgs}=120$~GeV \cite{DKMOR}. Likely, this assumption is not well
justified. In particular, the missing mass resolution and proton tagging efficiency may worsen at lower masses.},
of
$$ \sigma^{\rm CEDP}(pp\to p+(\bb)+p)\simeq 1.4\frac{\Delta M}
{1~{\rm GeV}}~{\rm pb}.$$ That is the signal-to-background ratio is only $S/B\sim 1/100$, and so even for an
integrated luminosity ${\cal L} = 300~{\rm fb}^{-1}$ for $\Delta M = 1$~GeV the significance of the signal is only
$3.7\sigma$. Here we have taken a $K$ factor of $K = 1.5$ for the QCD $\bb$ background, and again used the cuts
and efficiencies quoted in Ref.\cite{DKMOR}. Therefore, to identify a light Higgs, it is desirable to study a
decay mode other than $H_1\to\bb$. The next largest mode is $H_1\to\tau\tau$, with a branching fraction of about
0.07.

The dependence of the results on the mass of the $H_1$ Higgs boson is
illustrated in Table 1.   Clearly the cross section decreases with
increasing mass.  On the other hand the signal-to-background ratio
increases.   Therefore for the case $M_{H_1} = 50$~GeV we see a slightly
improved statistical significance of $4.4\sigma$ for the $\bb$ decay
mode.

\begin{table}[htb]
\begin{center}
\begin{tabular}{|l|c|c|c|c|}\hline

$M(H_1)$~GeV  &  cuts  &  30  &  40  &  50 \\ \hline

$\sigma(H_1){\rm Br}(\bb)$  &  $a$  & 45  & 14  & 6 \\
$\sigma^{\rm QCD}(\bb)$  &  $a$  &  16000  &  1400  &  200 \\
$A_{\bb}$  &     &  0.14  &  0.07  &  0.04 \\ \hline

$\sigma(H_1){\rm Br}(\tau\tau)$  &  $a,b$  & 1.9  &  0.6  & 0.3 \\
$\sigma^{\rm QED}(\tau\tau)$  &  $a,b$  &  0.2  &  0.1  &  0.04 \\
$A_{\tau\tau}$  &   $b$  &  0.2  &  0.1  &  0.05 \\ \hline\hline

$M(H_2)$~GeV  &    &  103.4  &  104.7  &  106.2 \\ \hline

$\sigma\dot {\rm Br} (H_2 \to 2H_1 \to 4b)$  &  $c$  &  0.5  &  0.5  &  0.5 \\
$\sigma\dot {\rm Br} (H_2 \to 2b)$  &  $a$  &  0.1  &  0.1  &  0.2 \\ \hline\hline

$M(H_3)$~GeV  &    &  141.9  &  143.6  &  146.0 \\ \hline

$\sigma\dot {\rm Br} (H_3 \to 2H_1 \to 4b)$  &  $c$  &  0.14  &  0.2  &  0.18 \\
$\sigma\dot {\rm Br}(H_3 \to 2b)$  &  $a$  &  0.04  &  0.07  &  0.1 \\ \hline
\end{tabular}
\end{center}
\caption{The cross sections (in fb) of the central {\em exclusive} diffractive production of $H_i$ neutral Higgs
bosons, together with those of the QCD($\bb$) and QED($\tau\tau$) backgrounds.  The acceptance cuts applied are
(a)~the polar angle cut $60^\circ<\theta(b~{\rm or}~\tau)<120^\circ$ in the Higgs rest frame,
(b)~$p_i^\perp>300$~MeV for the forward outgoing protons and (c)~the polar angle cut $45^\circ < \theta (b) <
135^\circ$. The azimuthal asymmetries $A_i$ are defined in eq.(12).}
\end{table}

\section{The $\tau\tau$ decay mode}

At the LHC energy, the expected cross section for exclusive diffractive
$H_1$ production, followed by $\tau\tau$ decay, is
\be \sigma\left(pp\to p+(H\to\tau\tau)+p\right) \sim 1.1~{\rm fb},
\label{eq:5} \ee
where the $60^\circ<\theta<120^\circ$ polar angle cut has already been included. Despite the low Higgs mass, we
note that the exclusive cross section is rather small.  As we already saw in (\ref{eq:2}), the cross section of
the hard subprocess $\hat\sigma(gg\to H)$ is approximately independent of $M_H$. Of course, we expect some
enhancement from the larger effective gluon--gluon luminosity ${\cal L}$ for smaller $M_H$. Indeed, it may be
approximated by ~\cite{INC,KKMRext}
\be {\cal L} \quad \propto \quad 1{\Big /}
(M_H + 16~{\rm GeV})^{3.3},
\label{eq:5a}
\ee
and gives an enhancement of about 18.8 (for $M_H=40$ GeV in comparison
with that for $M_H=120$ GeV).

\begin{figure}[htb]
\vspace{2ex} \centerline{\epsfxsize=3cm\epsfbox{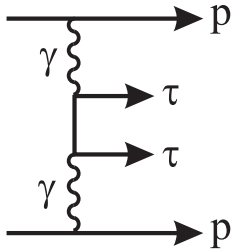}}
\caption{The QED background to the $H\to\tau\tau$ exclusive signal.}
\label{fig:2}
\end{figure}

On the other hand, in the appropriate region of SUSY parameter space, the
CP-even $H\to gg$ vertex, $g_S$, is almost 2 times smaller\cite{cox,Lee}
than that of a Standard Model Higgs, giving a
suppression of 4. Also the ratio $B(H\to \tau\tau)/B(H\to\bb)$ gives a
further suppression of about 12.
Although the $\tau\tau$ signal has the
advantage that there is practically no QCD background\footnote{ There may be background caused by
a pair of high $E_T$ ($\sim 15$ GeV) gluons being misidentified as a $\tau\tau$ pair.
To suppress such a background down to the level of $S/B\sim 1$,
the probability, $P_{g/\tau}$, that a gluon is misidentified as a $\tau$
must be less than about 1/750, assuming that the missing mass
resolution is $\Delta M=1$ GeV. In \cite{Zepp}, for an inclusive event, the
probability $P_{g/\tau}$ was evaluated as 1/500. Thus it seems reasonable
to suppose that the probability $P_{g/\tau}<1/750$ can be achieved in the much cleaner
environment of an exclusive diffractive (CEDP) event.},
exclusive $\tau^+\tau^-$
events may be produced by $\gamma\gamma$ fusion, see Fig.~\ref{fig:2}.
The cross section for this latter QED process is appreciable. It is enhanced by two large logarithms,
$\ln^2(t_{\rm min}R_p^2)$, arising from the integrations over the transverse momenta of the outgoing protons (that
is of the exchanged photons). The lower limit of the logarithmic integrals is given by
\be t_{\rm min} \ \simeq \ -(xm_p)^2 \ \simeq \ -\left(\frac{M_H}{\sqrt s}m_p\right)^2, \label{eq:6} \ee
while the upper limit is specified by the slope $R_p^2$ of the proton form factor. To suppress the QED background,
one may select events with relatively large transverse momenta of the outgoing protons. For example, if
$p_{1,2}^\perp > 300$~MeV, then the cross section for the QED
background, for $M_{\tau\tau}=40$ GeV,
is about\footnote{As we consider sizeable
$p_{1,2}^\perp$, we account for both the $F_1$ and $F_2$
electromagnetic proton form factors.}
\be \sigma_{\rm QED}(pp\to p + \tau\tau + p) \ \simeq \ 0.1
\frac{\Delta M}{1~{\rm GeV}}~{\rm fb},  \label{eq:7} \ee
while  the signal (\ref{eq:5}) contribution is diminished by the cuts,
$p^\perp_{1,2}>300$ MeV, down to 0.6 fb.  Thus, assuming an
experimental missing mass resolution of $\Delta M\sim 1$~GeV, we obtain
a healthy signal-to-background ratio of $S/B \sim 6$ for $M_{H_1} \sim
40$~GeV.

Note that in all the estimates given above, we include the appropriate soft survival factors $S^2$---that is the
probabilities that the rapidity gaps are not populated by the secondaries produced in the soft rescattering.  The
survival factors were calculated using the formalism of Ref.~\cite{KMRsoft}. Moreover, here we account for the
fact that only events with proton transverse momenta $p_{1,2}^\perp>300$~MeV were selected.  In particular, for
the QED process, we have $S^2\simeq 0.7$, rather than the value $S^2\simeq 0.9$, which would occur in the absence
of the cuts on the proton momenta\footnote{Without the momenta cuts, the main QED contribution comes from small
$p_{1,2}^\perp$, that is from large impact parameters $b^\perp\gg R_p$, where the probability of soft rescattering
is already small, see \cite{KMRphot} for details.}.

\section{ Azimuthal asymmetry of the outgoing protons}

A specific prediction, in the case of a CP-violating Higgs boson, is the asymmetry in the azimuthal $\varphi$
distribution of the outgoing protons, caused by the interference of the CP-odd and CP-even vertices, that is
between the two terms in (\ref{eq:1}). The polarisations of the incoming active gluons are aligned along their
respective transverse momenta, $Q_\perp - p_1^\perp$ and $Q_\perp + p_2^\perp$. Hence the contribution caused by
the second term, $g_P$, is proportional to the vector product
$$\vec{n}_0 \cdot (\vec{p}_1^\perp \times \vec{p}_2^\perp) \sim \sin\varphi,$$
where $\vec{n}_0$ is a unit vector in the beam direction, $\vec{p}_1$.
The sign of the angle $\varphi$ is fixed by the four-dimensional structure of the
second term in (\ref{eq:1}); see
\cite{KKMRCentr} for a detailed discussion. Of course, due to the P-even, $J_z=0$ selection rule,
 this (P-odd)
contribution is suppressed in the amplitude by
$p_1^\perp p_2^\perp/Q_\perp^2$, in comparison with that of the
P-even $g_S$ term. Note that there is a partial compensation of the suppression due to the ratio $g_P/g_S \sim 2$.
Also the soft survival factors $S^2$ are higher for the pseudoscalar and interference
terms, than for the scalar term.

An observation of the azimuthal
 asymmetry may therefore be a direct indication of the existence of
CP-violation (or P-violation in the case of CEDP) in the Higgs sector\footnote{In Ref.~\cite{CL} (see also
\cite{GG,KKSZ,GT}) a suggestion, along the same lines, was made for the explicit observation of CP-violating
effects. There, various polarization asymmetries in two-photon fusion Higgs production processes were discussed.
In the absence of absorptive effects, the azimuthal asymmetry $A$ may be expressed, via gluon helicity amplitudes,
in the same way as the quantity $A_2$ of \cite{CL}, written in terms of photon helicities.}. Neglecting the
absorptive effects (of soft rescattering), we find, for example, an asymmetry
\be A=\frac{\sigma(\varphi<\pi)-\sigma(\varphi>\pi)} {\sigma(\varphi<\pi)+\sigma(\varphi>\pi)} = 2{\rm Re}(g_S
g_P^*) r_{S/P} (2/\pi)/(|g_S|^2 + |r_{S/P} g_P|^2/2).
\ee
Here (numerically small) parameter $r_{S/P}$ reflects the suppression of the P-odd contribution due to the
selection rule discussed above.

At the LHC energy in the absence of rescattering effects $A\simeq 0.09$ for $M_{H_1}=40$ GeV.
 However we find soft rescattering tends to wash out the azimuthal distribution,
and to weaken the asymmetry. Besides this the real part of the
rescattering amplitude multiplied by the imaginary part of the
pseudoscalar vertex $g_P$ (with respect to $g_S$)
gives some negative contribution.
 So finally we predict $A\simeq 0.07$.
  For the lower Tevatron energy, the admixture of the P-odd
amplitude is larger, while the probability of soft rescattering is
smaller. Therefore, at $\sqrt s=2$ TeV, we find that asymmetry is twice
as large, $A\sim 0.17$. On the other hand the effective $gg^{PP}$
luminosity ${\cal L}$ and the corresponding cross section of $H_1$
(CEDP) production is 10 times smaller (for $M_{H_1}=40$ GeV).

The asymmetries expected at the LHC, with and without the cut $p_{1,2}^\perp>300$~MeV
on the outgoing protons, are shown for different $H_1$ masses in Table 1.
The asymmetry decreases with increasing Higgs mass, first, due to the
decrease of $|g_P|/|g_S|$ ratio in this mass range and, second, due to
the extra suppression of the P-odd amplitude arising from the factor
$p_1^\perp p_2^\perp/Q_\perp^2$ in which the typical value of $Q_\perp$
in the gluon loop increases with mass.

\section{ Heavy $H_2$ and $H_3$ Higgs production with $H_1H_1$ decay}

Another possibility to study the Higgs sector in the CPX scenario is to
observe central exclusive diffractive production (CEDP) of the heavy neutral $H_2$
and $H_3$ Higgs bosons, using the $H_2,H_3\to H_1 + H_1$ decay modes.
For the case we considered above ($\rm tan\beta=4$,
$\phi_{\rm CPX}=90^\circ$, $M_{H_1}=40$ GeV), the masses of the heavy bosons
bosons are $M_{H_2}=104.7$ GeV and $M_{H_3}=143.6$ GeV. At the LHC energy, the CEDP cross
sections of the $H_2$ and $H_3$ bosons are not too small --
$\sigma^{\rm CEDP}=1.5$ and $0.9\ {\rm fb}$ respectively. When the
branching fractions, Br$(H_2\to H_1H_1)=0.84$, Br$(H_3\to H_1H_1)=0.54$
and Br$(H_1\to\bb)=0.92,$ are included, we find
$$\sigma(pp\to p+(H\to\bb\ \bb)+p)=1.1\ {\rm and}\ 0.4\ {\rm fb}$$
for $H_2$ and $H_3$ respectively. Thus there is a chance
to observe, and to identify,
the central exclusive diffractive production of all three neutral
Higgs bosons, $H_1, H_2$ and $H_3$, at the LHC.

The QCD background for exclusive diffractive production of four $b$-jets is
significantly less than the signal.   Other decay channels are also worth
mentioning.  For a very light boson, say $M_{H_1} = 30$~GeV, it is also possible
to produce four $b$-jets via the cascade $H_3\to H_2H_1\to 4b$-jets. However,
the expected cross section is about 0.02 fb, which looks too low to be useful.
A larger cross section is expected for the direct $H_2\to\bb$ decay, where
the branching fraction Br$(H_2\to\bb)=0.14$ for $M_{H_1}=40$ GeV leads to
the cross section
$\sigma(p+(H_2\to\bb)+p)$ = 0.2 fb.  Note that in this case, we only need to
tag two, and not four, $b$-jets.  So the detection efficiency is about a
factor of 1/0.6 larger.  The situation is even better for $M_{H_1} = 50$~GeV,
where Br$(H_2\to\bb)=0.25$ and
$\sigma(p+(H_2\to\bb)+p)$ = 0.4 fb.  If it is possible to compare the $4b$-
and $2b$-jet signals, then it will allow a probe of the nature of the $H_2$ boson.
Finally, for the heaviest boson, $H_3$, the decay mode $H_3\to H_1+Z$ is not small,
with a branching fraction of Br$(H_3\to H_1+Z)=0.27$ for $M_{H_1}=40$ GeV.

\section{Central Higgs production with double diffractive dissociation}

To enhance the Higgs signal we study a less exclusive reaction than $pp\to p + H + p$,
and allow both of the incoming protons to dissociate.  In Ref.\cite{INC} it was
called double diffractive {\it inclusive} production, and was written
\be pp\to X + H + Y.  \label{eq:CIDP} \ee
Now there is no form factor suppression as the initial protons are destroyed.
Also the cross section is larger due to the increased $p_i^\perp$ phase space.
Moreover the cross section is also enhanced because we no longer
have the P-even selection rule, and so the pseudoscalar $gg\to H$ coupling, $g_P$,
becomes active.  The cross section for inclusive production, via
central double dissociation (CDD) process, is found by using (i) the effective $gg^{PP}$
luminosity of Ref.\cite{INC}, (ii) the probability, $S^2$, that the gaps survive
soft rescattering, calculated using model II of \cite{KKMR}, and (iii) the
opacity of the proton given in \cite{KMRsoft}.  Typical results, for
the LHC energy, are shown in Table 2.  For the Tevatron energy, the cross section
appears too small, and even for a light boson of mass $M_{H_1}=30$ GeV we have
Br$(H_1\to \tau\tau)\sigma<1.5$fb, while the QED background is about 15 fb.

\begin{table}[htb]
\begin{center}
\begin{tabular}{|l|c|c|c|}\hline

$M(H_1)$~GeV  &  30  &  40  &  50 \\ \hline

$\sigma(H_1){\rm Br}(\tau\tau)$  &  19 (4)  &  6 (2)  & 2.6 (0.8) \\
$\sigma^{\rm QED}(\tau\tau)$    &  66 (2.2)  &  30 (1.5)  &  15 (0.9) \\ \hline\hline

$M(H_2)$~GeV   &  103.4  &  104.7  &  106.2 \\ \hline

$\sigma\dot {\rm Br} (H_2 \to 2H_1 \to 4b)$   &  4 (2)  &  4 (2)  &  3.5 (2) \\ \hline\hline

$M(H_3)$~GeV  &   141.9  &  143.6  &  146.0 \\ \hline

$\sigma\dot {\rm Br} (H_3 \to 2H_1 \to 4b)$   &  1.5 (0.8)  &  2.2 (1.2)  &  2 (1.1) \\ \hline
\end{tabular}
\end{center}
\caption{The cross sections (in fb) for the central production of $H_i$ neutral Higgs bosons by {\em inclusive}
double diffractive dissociation, together with that of the QED($\tau\tau$) background.  A polar angle acceptance
cuts of $60^\circ<\theta(b~{\rm or}~\tau)<120^\circ$
 ($45^\circ<\theta(b)<145^\circ$)
in the Higgs rest frame is applied for the case of $H_1$  ($H_2,H_3$) bosons.
 The numbers in brackets
correspond to the imposition of the additional cut of $E^\perp_i>7$~GeV for the proton dissociated systems.}
\end{table}

Of course, the missing mass method cannot be used to measure the mass of the Higgs boson for central production
with double dissociation
(CDD). Therefore the mass resolution will be not so good as for CEDP; we evaluate the background for $\Delta M$ =
10 GeV.  Moreover, with the absence of the $J_z=0$ selection rule, the LO QCD $\bb$-background is not suppressed.
Hence we study only the $\tau\tau$ decay mode for the light boson, $H_1$, and the four $b$-jet final state for the
heavy $H_2$ and $H_3$ bosons.

The background to the $H_1\to\tau\tau$ signal arises from the $\gamma\gamma\to\tau\tau$
QED process.  It is evaluated in the equivalent photon approximation.  The photon flux,
\be
N_\gamma = \frac{\alpha}{\pi}
\frac{dq^2}{q^2}\frac{dx}{x} F_2(x,q^2),
\label{eq:35a}
\ee
was calculated using LO MRST2001 partons\cite{MRST01}, with the integral over
the photon transverse momentum running from $q=m_\rho$ up to $q=M_{\tau\tau}/2$.
The lower limit is approximately where the $\gamma^*p$ cross section becomes
flat and loses its $\sigma(\gamma^*p) \sim 1/q^2$ behaviour.  The upper limit reflects the
dependence of the $\gamma\gamma\to\tau\tau$ matrix element on the virtuality of
the photon.  From Table 2 we see that the $H_1$ signal for inclusive
diffractive production, (\ref{eq:CIDP}), exceeds the exclusive signal by more than
a factor of ten.  On the other hand the signal-to-background ratio is worse;
$S/B_{QED}$ is about 1/5.  Moreover there could be a huge background due the
misidentification of a gluon dijet as a $\tau\tau$-system.  To make this QCD
background satisfy $B_{QCD}<S$, would require the probability of misidentifying a gluon
as a $\tau$ lepton to be $P_{g/\tau}<1/1500$.

For the four $b$-jet signals of the heavy $H_2$ and $H_3$ bosons, the QCD background can be suppressed by
requiring each of the four $b$-jets to have polar angle in the interval $(45^\circ,135^\circ)$, in the frame where
the four $b$-jet system has zero rapidity. However in the absence of a good mass resolution, that is with
only\footnote{However this resolution is still sufficient to separate the $H_2$ and $H_3$ bosons.} $\Delta M=10$
GeV, we expect the four $b$-jet background to be 3-5 times the signal.  Nevertheless these signals are still
feasible, with cross sections of the order of a few fb.   For example, with an integrated luminosity of ${\cal L}
= 300~{\rm fb}^{-1}$ and an efficiency of $4b$-tagging of $(0.6)^2$ \cite{DKMOR}, we predict about 400 $H_2$
events and 200 $H_3$ events.  Taking the background-to-signal ratio to be $B/S =4$, we then have a statistical
significance of about $10\sigma$ for $H_2$ and $6\sigma$ for $H_3$.

 \begin{figure}[htb]
\centerline{\epsfxsize=6cm\epsfbox{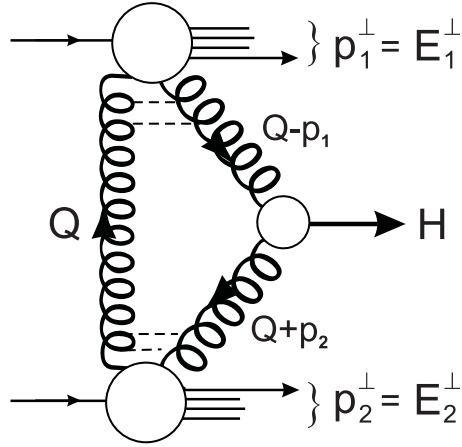}}%
\caption{Central Higgs production with double diffractive dissociation (CDD), in
which the incoming protons dissociate into systems with transverse energies $E^\perp_i$}
\label{fig:3}
\end{figure}

The inclusive CDD kinematics allow a study of CP-violation, and the separation of
the contributions coming from the scalar and pseudoscalar $gg\to H$ couplings,
$g_S$ and $g_P$ of (\ref{eq:1}), respectively.  Indeed, the polarizations of the incoming active
gluons are aligned along their transverse momenta, $\vec{Q}_\perp-\vec{p}^\perp_1$ and
$\vec{Q}_\perp+\vec{p}^\perp_2$.  Hence the $gg\to H$ fusion vertices take the forms
\be
V_S = (\vec{Q}_\perp-\vec{p}^\perp_1) \cdot (\vec{Q}_\perp+\vec{p}^\perp_2) g_S
\ee
\be
V_P = \vec{n}_0 \cdot [(\vec{Q}_\perp-\vec{p}^\perp_1) \times (\vec{Q}_\perp+\vec{p}^\perp_2)] g_P,
\ee
where $g_S$ and $g_P$ are defined in (\ref{eq:1}).

For the exclusive (CEDP) process the momenta $p^\perp_{1,2}$ were limited by the
proton form factor, and typically $Q^2\gg p^2_{1,2}$.  Thus
\be
V_S =  g_S~Q^2_\perp  \;\;\;\;\; {\rm while} \;\;\;\;\;
  V_P = g_P~(\vec{n}_0\cdot[\vec{p}^\perp_2\times \vec{p}^\perp_1]).
\ee
On the contrary, for double diffractive dissociation production (CDD) $Q^2 < p^2_{1,2}$.
In this case
\be
V_S =  g_S~p^\perp_1 p^\perp_2 {\rm cos}\varphi  \;\;\;\;\; {\rm and} \;\;\;\;\;
  V_P = g_P~p^\perp_1 p^\perp_2 {\rm sin}\varphi.
\ee
Moreover we can select events with large outgoing transverse momenta of
the dissociating systems, say $p^\perp_{1,2}> 7$ GeV, in order to
make reasonable measurements of the directions of the vectors
$\vec{p}_1^\perp=\vec{E}^\perp_1$ and  $\vec{p}_2^\perp=\vec{E}^\perp_2$.
Here $E^\perp_{1,2}$ are the transverse energy flows of the dissociating
systems of the incoming protons.  At LO, this transverse energy is carried
mainly by the jet with minimal rapidity in the overall centre-of-mass
frame.  The azimuthal angular distribution has the form\footnote{In the CP-conserving
case, an idea similar in spirit was considered in Ref.\cite{DKOSZ}, where it was suggested
to measure the azimuthal correlations of the two tagged jets in inclusive Higgs
production.  However the proof of the feasibility of such an approach in non-diffractive processes
requires further detailed studies of the possible dilution of the effect due to the
parton showers in the inclusive environment of the jets.}
\be
\frac{d\sigma}{d\varphi} = \sigma_0 (1+ a~{\rm sin}2\varphi + b~{\rm cos}2\varphi),
\ee
where the coefficients are given by
\be
a=\frac{2{\rm Re}(g_Sg_P^*)}{|g_S|^2+|g_P|^2} \;\;\;\;\; {\rm and} \;\;\;\;\;
b=\frac{|g_S|^2-|g_P|^2}{|g_S|^2+|g_P|^2}.
\ee
Note that the coefficient $a$ arises from scalar-pseudoscalar interference, and reflects the presence of a T-odd
effect. Its observation would signal an explicit CP-violating mixing in the Higgs sector. On the other hand, in
the absence of CP-violation,the sign of the coefficient b reveals the CP-parity of the new boson\footnote{Note
that we may search for any new pseudoscalar boson produced by the CDD process by looking for the corresponding
azimuthal distribution, $d\sigma/d\varphi \sim {\rm sin}^2\varphi$.}.

The predictions for the coefficients are given in Table 3 for different values of the
Higgs mass, namely $M_{H_1}$ = 30, 40 and 50 GeV.  The coefficients are of appreciable size and, given
sufficient luminosity,  may be measured at the LHC.   Imposing the cuts $E^\perp_i > 7$
GeV reduces the cross sections by about a factor of two, but does not alter the
signal-to-background ratio,
$S/B_{QCD}$.  However the cuts do give increased suppression of the QED
$\tau\tau$ background and now, for the light $H_1$ boson, the ratio $S/B_{QED}$
exceeds one.  We emphasize here that, since we have relatively large $E^\perp$,
the angular dependences are quite insensitive to the soft rescattering corrections.

\begin{table}[htb]
\begin{center}
\begin{tabular}{|c|rr|rr|rr|}\hline

$M(H_1)$~GeV  & \multicolumn{2}{|c|}{30}  &  \multicolumn{2}{|c|}{40}  &  \multicolumn{2}{|c|}{50} \\ \hline

& \multicolumn{1}{|c}{$a$}  &  \multicolumn{1}{c|}{$b$}  &  \multicolumn{1}{|c}{$a$}  &  \multicolumn{1}{c|}{$b$}
&  \multicolumn{1}{|c}{$a$}  & \multicolumn{1}{c|}{$b$}   \\ \hline

$H_1$  &  $-0.53$  &  $-0.73$  &  $-0.56$  &  $-0.55$  &  $-0.53$  &  $-0.33$ \\
$H_2$  &  0.44  &  0.90  &  0.41  &  0.91   &  0.37  &  0.92 \\
$H_3$  &  $-0.38$  &  0.92  &  $-0.40$  &  0.91  &  $-0.42$  &  0.90 \\
\hline
\end{tabular}
\end{center}
\caption{The coefficients in the azimuthal
distribution $d\sigma/d\varphi = \sigma_0 (1+ a\sin 2\varphi + b \cos 2\varphi)$,
where $\varphi$ is the azimuthal angle between the $E^\perp$ flows of the two proton dissociated systems.
If there were no CP-violation, then the coefficients would be $a=0$ and $|b|=1$.}
\end{table}

\section{Conclusions}

We have evaluated the cross sections, and the corresponding backgrounds, for
the central double-diffractive production of the (three neutral) CP-violating
Higgs bosons at the LHC.  This scenario is of interest since
even a very light boson of mass about 30 GeV is not experimentally ruled
out for some range of the MSSM parameters.

We have studied the production of the three states, $H_1, H_2, H_3$, both
with exclusive kinematics, $pp\to p + H + p$ which we denoted CEDP,
and in double-diffractive reactions
where both the incoming protons may be destroyed, $pp\to X + H + Y$
which we denoted CDD.  Recall that a
+ sign denotes the presence of a large rapidity gap.  Proton taggers are required
in the former processes, but not in the latter.  Typical results are summarised
in Tables 1 and 2, respectively.  The cross sections are not large, but should be accessible
at the LHC.  The uncertainties in the calculation of the exclusive cross sections
were discussed in Refs.\cite{KKMRCentr,KKMRext}.  For the light $H_1$ boson, where
the contribution from the low $Q_\perp$ region is more important, the uncertainty
is much larger.  Recall that for the semi-inclusive CDD processes the effective
gluon-gluon ($gg^{PP}$) luminosity is calculated using the LO formula.  Thus we
cannot exclude rather large NLO corrections.  On the other hand, for CDD, the
values of the cross sections are practically
insensitive to the contributions from the infrared domain.  Moreover, with the {\it skewed}
CDD kinematics, the NLO BFKL corrections are expected to be much smaller
than in the forward (CEDP) case.  So we may expect an uncertainty of the
predictions to be about a factor of 3 to 4, or even better.

It would be very informative to measure the azimuthal angular dependence
of the outgoing proton systems, for both the CEDP and CDD processes.  Such measurements
would reveal explicitly any CP-violating effect, via the interference of
the scalar and pseudoscalar $gg\to H$ vertices.

Finally, we recall the advantages of diffractive, as compared to the
non-diffractive, production of Higgs bosons:

i) a much better mass Higgs resolution is obtained by the
missing mass method for exclusive events,

ii) a clean environment, which may be important
to identify four $b$-jets with transverse
momenta $p_T\sim M_{H_1}/2\sim 20$ GeV (for the
non-diffractive process, at the LHC energy, the QCD backgroud  may be too
large),

iii) a possibility to measure  CP-property of the Higgs boson and to detect CP-violation (note that the
asymmetries $A_{\bb}$ and $A_{\tau\tau}$ are explicit manifestations of CP violation at the
 quark level),

Next, assuming that P and C parities are conserved,

iv) the existence of the P-even, $J_z=0$ selection rule for LO central exclusive diffractive production, which
means that we observe an object of natural parity (most probably $0^+$); the analysis of the azimuthal angular
distribution of the outgoing protons may give additional information about the spin of the centrally produced
object~\cite{KKMRCentr},

v) in addition we know that an object produced by the diffractive process (that is by Pomeron-Pomeron fusion) has
positive C-parity, is an isoscalar and a colour singlet\footnote{An instructive topical example, which illustrates
the power of CEDP as a spin-parity analyser, concerns the determination of the quantum numbers of the recently
discovered $X(3872)$ resonance\cite{chi3872}. A knowledge of its C-parity is important to understand its nature.
If it is a C = +1 state with spin-parity $0^+$ or $2^+$ then it may be even seen in CDD production with a large
rapidity gap on either side of its J/$\psi~\pi^+\pi^-$ decay.  Forward proton tagging would, of course, allow a
better spin-parity analysis.}.

The properties listed above should help to distinguish the $H_2$ and $H_3$ four-jet decay channels from the production of a
SUSY particle, followed by a `cascade'-like decay.

%
%
%
%

\section*{Acknowledgements}

We thank Jeff Forshaw, Risto Orava, Albert de Roeck, Sasha Nikitenko,
Apostolos Pilaftsis and, especially, Brian Cox
and Jae Sik Lee for useful discussions.
ADM thanks the Leverhulme trust for an Emeritus Fellowship and MGR thanks the IPPP at the University of
Durham for hospitality. This work was supported by
the UK Particle Physics and Astronomy Research Council, by grant RFBR 04-02-16073
and by the Federal Program of the Russian Ministry of Industry, Science and Technology
SS-1124.2003.2.

\end{document}